\documentclass[prd,aps,preprint,epsfig,floats]{revtex4}
\usepackage{graphics}
\usepackage{epsfig}
\usepackage{graphicx}

\begin{document}

%\draft
\preprint{\vbox{
\hbox{UMD-PP-06-002} }}
\title{\Large\bf Grand unification of $\mu-\tau$ Symmetry}
\author{\bf R.N. Mohapatra, S. Nasri and Hai-Bo Yu }

\affiliation{ Department of Physics, University of Maryland, College
Park, MD 20742, USA}

\date{February, 2006}
%\maketitle

\begin{abstract}
Near maximal neutrino mixing needed to understand atmospheric
neutrino data can be interpreted to be a consequence of an
interchange symmetry between the muon and tau neutrinos in the neutrino
mass matrix in the flavor basis. This idea
 can be tested by a measurement of the neutrino mixing
parameter $\theta_{13}$ and looking for its correlation with
$\theta_{23}-\pi/4$. We present a supersymmetric SU(5) grand unified
model for quarks and leptons which obeys this exchange symmetry and
is a realistic model that can fit all observations. GUT embedding
shifts $\theta_{13}$ from its zero $\mu-\tau$ symmetric value to a nonzero 
value keeping it under an upper limit.
 \end{abstract}
\maketitle
\section{Introduction}
Observation of nonzero neutrino masses and determination of two of
their three mixing parameters by experiments have raised the hope
that neutrinos may provide a clue to flavor structure among
quarks\cite{rev}. In order to make progress in this direction
however, one needs knowledge of the detailed nature of the quark-lepton
connection e.g. whether there is an energy scale
where quarks and leptons are unified into one matter (or grand
unification of matter). While there are similarities between quarks
and leptons that make such an unification plausible, there are also
many differences between them which may apriori point the other way:
for instance, the mixing pattern
among quarks is very different from that among leptons and the neutrino
mass matrices in the flavor basis exhibit symmetries for
which there apparently is
no trace among quarks. Two examples of such apparent
lepton-exclusive symmetries are : (a) discrete $\mu-\tau$
symmetry\cite{mutau,moh} of the neutrino mass matrix in the flavor
basis indicated by maximal atmospheric mixing angle and small
$\theta_{13}$ and (b) continuous $L_e-L_\mu-L_\tau$\cite{emutau}
symmetry, which will be indicated if the mass hierarchy among
neutrinos is inverted.

If neutrinos are
Majorana fermions, they are likely to acquire masses from very different
mechanisms e.g. one of the various seesaw mechanisms
which involve completely independent flavor structure (say for example
from right handed neutrinos) than quarks. The apparent disparate pattern
for quark and leptons mixings then need not argue against
eventual quark-lepton unification.
In fact there are now many grand unification models (where quarks and
leptons are unified at short distances) where small quark mixings and
large lepton mixings along with all their masses can be understood with
very few assumptions in a seesaw framework\cite{rev}.

In this paper we address the question as to whether there could be
an apparently pure leptonic symmetry such as $\mu-\tau$ symmetry in
the neutrino mass matrix in the flavor basis (i.e. the basis where
charged leptons are mass eigenstates), which is part of a general
family symmetry within a quark-lepton unified framework such as a
grand unified model. We particularly focus on this symmetry since
there appears to be some hint in favor of this from the present
mixing data. In the exact symmetry limit, the mixing parameter
$\theta_{13}=0$\cite{mutau} and breaking of the symmetry not only
implies a small nonzero value for $\theta_{13}$ but also leads to a
correlation between $\theta_{13}$ with $\theta_{23}-\pi/4$, which
can be used to test for this idea\cite{moh}. This question has been
discussed at a phenomenological level in several recent
papers\cite{nishiura} but to the best of our knowledge no
full-fledged gauge model has been constructed. Indeed most gauge
models for $\nu_\mu-\nu_\tau$ symmetry discussed in the literature
treat leptons separately from quarks\cite{grimus}.

One simple way to have quark flavor structure completely separated
from that of leptons and yet have quark-lepton unification is to
use the double seesaw\cite{valle} framework where neutrino flavor
texture from ``hidden sector'' singlet fermions (e.g. SO(10)
singlets ) which are completely unrelated to quarks (for examples
of such models, see \cite{fram,lind}). One can then have any pure
``leptonic'' symmetry on the hidden singlets without at the same
time interfering with quark flavor texture. A necessary feature of
such models is that one must introduce new fermions into the
model. A question therefore remains as to whether one could do
this without expanding the matter sector. In this paper, we
propose such an
  approach without introducing new fermions within a realistic SU(5) GUT
framework that unifies quarks and leptons. We demand the full theory
    prior to symmetry breaking to obey a
symmetry between the second and third generation (or a generalized
version of ``$\mu-\tau$'' symmetry). The neutrino masses are
assumed to arise from a triplet seesaw (type II)\cite{triplet}
mechanism, which disentangles the neutrino flavor structure from
the quark flavor structure. The quark mass matrices are however
constrained by the $\mu-\tau$ symmetry. The quark mixing angles
then introduce departures from exact $\mu-\tau$ symmetry results
and lead to nonzero $\theta_{13}$ as well as departures from
maximal atmospheric mixing.

The model consists of a minimal set of Higgs bosons which are
anyway required to reconcile the charged fermion masses in the
minimal SU(5) model. We find that the requirement of $\mu-\tau$
symmetry for neutrinos can be imposed on the model without
contradicting observed charged fermion masses and mixings. As
noted, the model predicts a nonzero value for $\theta_{13}$
correlated with the departure of $\theta_{23}$ from its maximal
value.

This paper is organized as follows: in sec. 2, we present the SU(5)
model with $\mu-\tau$ symmetry; in sec. 3, we discuss coupling
unification in the model since we have a new scale around $10^{14}$
GeV to implement the type II seesaw for neutrino masses. We close
with concluding remarks in sec. 4.

\section{SUSY SU(5) Model with $\mu-\tau$ symmetry}
As in the usual SU(5) model, matter fields are assigned to $\bar{\bf
5}\equiv F_\alpha$ and ${\bf 10}\equiv T_\alpha$ (with $\alpha=
1,2,3$ denotes the generation index). We choose the Higgs fields to
belong to the multiplets {\bf 24} (denoted by $\Phi$ and used to
break the SU(5) symmetry down to the standard model); ${\bf
5}\oplus\bar{\bf 5}$ (denoted by $h+\bar{h}$) and ${\bf
45}\oplus\bar{\bf 45}$ (denoted by $H+\bar{H}$) used to give masses
to fermions) and ${\bf 15}\oplus\bar{\bf 15}$ (denoted by
$S+\bar{S}$) to give masses to neutrinos via the type II seesaw
mechanism\cite{triplet}.

The matter and Higgs fields transform under the $\mu-\tau$ discrete
flavor symmetry as follows:
\begin{eqnarray}
\nonumber F_\mu &\leftrightarrow& F_\tau \\ \nonumber (h,\bar{h})
&\leftrightarrow& (h, \bar{h}) \\  (H,\bar{H}) &\leftrightarrow&
(-H, -\bar{H})
\end{eqnarray}
and all other fields are singlets under this transformation. In
this model, the matter part of the superpotential can be written as
\begin{eqnarray}
W=Y_{15}FFS+Y_{5}TTh+Y_{\bar{5}}TF\bar{h}+Y_{45}TFH.
\end{eqnarray}

After the electro-weak symmetry breaking, the mass matrices for the
standard model fermions are given by

\begin{eqnarray}
M_\nu &=& Y_{15}<S>=\left[\matrix{X&Y&Y\cr Y&Z&W \cr Y&W&Z}\right]\\
M_u &=& Y_{5}<h>=\left[\matrix{A&B&C\cr B&D&E \cr C&E&F}\right]\\
M_d &=& Y_{\bar5}<\bar h>+Y_{45}<H>=\left[\matrix{A_1&B_1&C_1\cr
E_1&D_1&F_1\cr E_1&D_1&F_1}\right]+\left[\matrix{0&0&0\cr
E_2&D_2&F_2\cr
-E_2&-D_2&-F_2}\right]\\
M_e &=& Y^T_{\bar5}<\bar
h>-3Y^T_{45}<H>=\left[\matrix{A_1&E_1&E_1\cr B_1&D_1&D_1\cr
C_1&F_1&F_1}\right]-3\left[\matrix{0&E_2&-E_2\cr 0&D_2&-D_2\cr
0&F_2&-F_2}\right]
\end{eqnarray}
where the various parameters characterising the mass matrices are
given in terms of the Yukawa couplings and vacuum expectation values
of fields as follows:$<S>,<h>,<\bar h>,<H>$ are vevs of $S,h,\bar
h,H$ respectively.

The mass matrices depend on nineteen parameters if we ignore CP
phases and there are seventeen experimental inputs (6 quark masses,
3 charged lepton masses, two neutrino mass difference squares plus
five mixing angles values and an upper limit on $\theta_{13}$). For
the sake of comparison, we note that if we generated neutrino masses
in the standard model using a Higgs triplet field, there would be 18
parameters in the absence of CP violation (9 from the quark sector,
3 from the charged lepton mass matrix and six from the neutrino
sector). When one embeds the standard model into a GUT SU(5), to be 
realistic, one needs to introduce {\bf 45} Higgs and its associated 
Yukawa couplings. In this case, the total number of parameters in the 
Yukawa sector is 24. 
In our model the requirement of $\mu-\tau$
symmetry has first led to a reduction in the total number by three and
furthermore grand unification has strongly correlated the down quark and 
charged lepton mass matrix, as expected. It is therefore not obvious that 
the model will be consistent with known data on fermion masses.

 To see if the model is phenomenologically acceptable,
 we first fit the masses of the charged leptons and down type quarks using
the mass values of leptons and quarks at GUT scale given in Ref.
\cite{dasparida}:
\begin{center}
\begin{tabular}{|c||c|}\hline
input observable & $tan\beta=10$ \\ \hline $m_u$ (MeV) &
$0.7238^{+0.1365}_{-0.1467}$ \\\hline $m_c$ (MeV) &
$210.3273^{+19.0036}_{-21.2264}$
\\ \hline $m_t$ (GeV) & $82.4333^{+30.2676}_{-14.7686}$ \\\hline$m_d$ (MeV) &
$1.5036^{+0.4235}_{-0.2304}$
\\\hline
$m_s$ (MeV) & $29.9454^{+4.3001}_{-4.5444}$
\\\hline
$m_b$ (GeV) & $1.0636^{+0.1414}_{-0.0865}$
\\\hline
$m_e$ (MeV) & $0.3585^{+0.0003}_{-0.0003}$  \\\hline $m_{\mu}$ (MeV)
& $75.6715^{+0.0578}_{-0.0501}$ \\\hline $m_{\tau}$ (GeV) &
$1.2922^{+0.0013}_{-0.0012}$\\ \hline
\end{tabular}
\end{center}

The values of parameters in the model are found by scanning the whole
parameter space under the constraint that we satisfy the
current experiment requirements of  $\theta_{13}$ and $\theta_{23}$. Note
that since in this model, neutrino mass matrix in the flavor basis is
$\mu-\tau$ symmetric, it is diagonalized by the matrix:
\begin{eqnarray}
U_\nu=\frac{1}{\sqrt2}
\left[\matrix{\sqrt2\cos\theta_\nu&\sqrt2\sin\theta_\nu&0\cr
-\sin\theta_\nu&\cos\theta_\nu&1\cr-\sin\theta_\nu&\cos\theta_\nu&-1}\right],
\end{eqnarray}
where $\theta_\nu$ is the solar mixing angle. The deviations of
$\theta_{13}$ and $\theta_{23}$ from $0$ and $\frac{\pi}{4}$
respectively should come from left-handed charged leptons mixing
matrix. Since these deviations have upper bounds, this puts an
nontrivial constraint on the charged lepton mass matrix of the
model; but since the charged lepton mass matrix is already
constrained by $\mu-\tau$ symmetry, it is nontrivial to get all
masses and mixings to fit. It turns out that the fitting for the
masses of leptons and quarks does not provide any bound on
$\theta_{23}$, however it gives quite stringent bound on
$\theta_{13}$.
 Using the relation $U_{MNS}=U_l^\dag
U_\nu$, one can write $\sin\theta_{13}$ and $\tan\theta_{23}$ as
\begin{eqnarray}
\sin\theta_{13}=\frac{1}{\sqrt2}|U_{l21}-U_{l31}|\\
\tan\theta_{23}=|\frac{U_{l22}-U_{l32}}{U_{l23}-U_{l33}}|
\end{eqnarray}
The $3\sigma$
experimental bounds of $\theta_{13}$ and $\theta_{23}$ are \cite{valle1}
\begin{eqnarray}
0.34\leq\sin^2\theta_{23}\leq0.68\\
\sin^2\theta_{13}\leq 0.051.
\end{eqnarray}
The scatter plot in Fig.1 gives $\sin^2\theta_{13}$ as a function of
sin$^2\theta_{23}$ allowing for 3 $\sigma$ uncertainty in all masses
except $m_e$ (chosen to be $0.3-0.4$ MeV), $m_\mu$ (chosen to be
$73-76$ MeV) and $m_d$ left free and $\theta_{23}$ within 3
$\sigma$.

\begin{figure}[h!]
\includegraphics[scale=0.8]{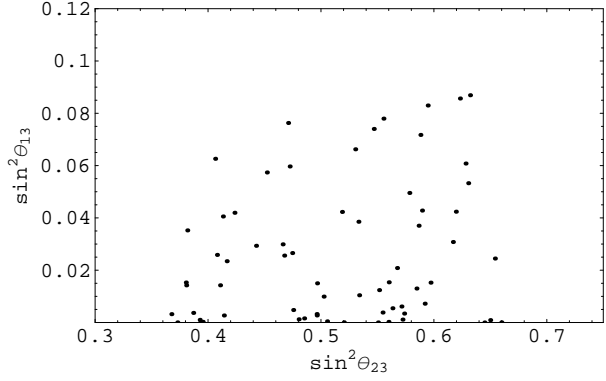}
\caption{Scatter  plot in the $\sin^2\theta_{23}$ and
$\sin^2\theta_{13}$ plane.}
\end{figure}

\begin{figure}[h!]
\includegraphics[scale=0.8]{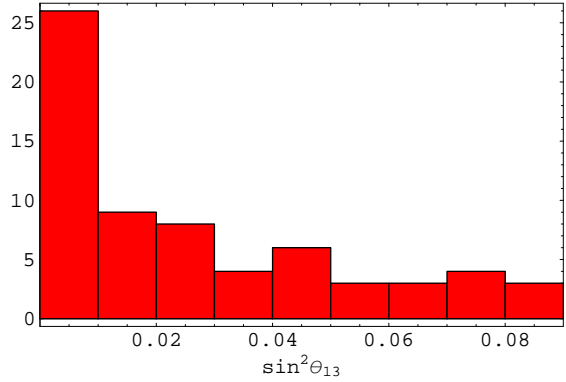}
\caption{Value distribution of $\sin^2\theta_{13}$. 67 percent
of fitting points have $\sin^2\theta_{13}\leq 0.03$ and 80 percent
have $\sin^2\theta_{13}\leq 0.05$.}
\end{figure}

Here, we give two typical fitting points for our model:

\noindent{\it (i) Case 1:}
\begin{eqnarray}
m_d &=& 0.355117\  {\rm MeV}\quad m_s=34.0438\ {\rm MeV} \quad
m_b=985.857\
{\rm MeV} \\
m_e &=& 0.356047\ {\rm MeV}\quad m_\mu=75.1597\ {\rm MeV}\quad
m_\tau=1336.14\ {\rm MeV}
\end{eqnarray}

\begin{eqnarray}
U_l=\left[\matrix{0.999327&0.036688&0.0000316411\cr
0.0366849&-0.999231&-0.0138381\cr
0.000476075&-0.01383&0.999904}\right]%\label{Ul}
\end{eqnarray}
For this case, we predict the following values for the neutrino mixing
parameters $\theta_{13}$ and $\theta_{23}$:
\begin{eqnarray}
\theta_{13}\simeq 0.026\\
\theta_{23}\simeq 44.3^\circ
\end{eqnarray}

\noindent{\it (ii) Case 2:}

\begin{eqnarray}
m_d &=& 0.336552\ {\rm MeV}\quad m_s=38.4364\ {\rm MeV} \quad
m_b=926.78\
{\rm MeV} \\
m_e &=& 0.381779\ {\rm MeV}\quad m_\mu=73.112\ {\rm MeV}\quad
m_\tau=1288.52\ {\rm MeV}
\end{eqnarray}

\begin{eqnarray}
U_l=\left[\matrix{0.959961&0.280133&0.000326329\cr
0.279872&-0.959014&-0.0443148\cr
0.0121011&-0.0426319&0.999018}\right]\label{Ul}
\end{eqnarray}
giving us
\begin{eqnarray}
\theta_{13}\simeq 0.19788\\
\theta_{23}\simeq 41.2^\circ
\end{eqnarray}
We therefore note that the value of the most probable value
for $\theta_{13}$ is in the range from $0.02-0.19$ with (as indicated in
Fig. 2) values below $0.1$ being much more probable.

Note that mass $m_d$ in both cases has almost same magnitude as
$m_e$ and is smaller than the central value at the GUT scale by
about $\sim \rm{1MeV}$. The reason for this is that $H$ is
$\mu-\tau$ odd, leading to
 zero entries in the $M_e,M_d$. Note that
in this model, we also have additional threshold correction from the
exchange of the gauginos, which make larger contribution to quarks
relative to the charged leptons of the corresponding generation due to
strong coupling of the gluinos.
 In particular, the gluino contribution to the
 tree level masses of the quarks can be significant if the
assumption of
proportionality between the A-terms and the Yukawa couplings is
abandoned. Fig. 3 gives a typical Feynman diagram contributing to
the quark masses\cite{raby}. The generic contribution to the $(i,j)$
element of the down quark mass matrix is given by:

%%%%%%%%%%%%%%%%%%%%%
\begin{figure}[h!]
\includegraphics[scale=1.5]{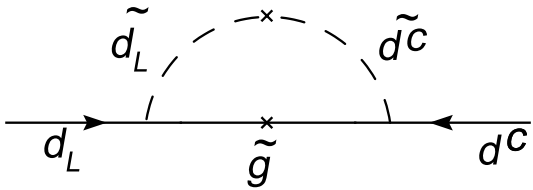}
\caption{One-loop SUSY threshold correction to $d$ quark mass due to 
gluino-squark exchange.}
\end{figure}

\begin{eqnarray}
\delta m_{d, ij}\simeq
\frac{2\alpha_s}{3\pi}\frac{M_{\tilde{g}}}{m^2_{\tilde{q}}}
(m^0_{d,ij}\mu\tan\beta+A^{(d)}_{ij}m_0)
\end{eqnarray}

Including this radiative correction only in the $11$ element of the down
quark mass matrix, one can get the down quark mass to be in agreement with
observations.
 We also note that the process of fitting the charged lepton and
down quark masses gives a definite rotation matrix that diagonalizes the
down quark mass matrix and contributes to the $V_{CKM}$. We then
appropriately choose the parameters in the symmetric up-quark mass matrix
so that we get the correct $V_{CKM}$.

\section{Other comments on the model:}

\subsection{Gauge coupling unification}
\begin{figure}[h]
\includegraphics[scale=1]{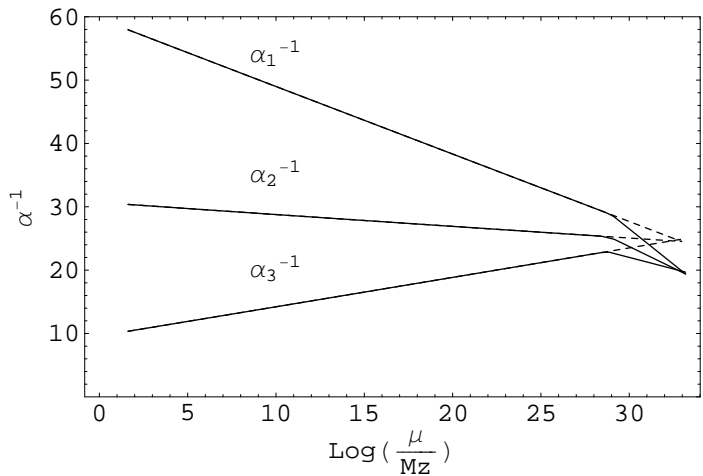}
\caption{Unification of the gauge couplings at two-loop level for
central values of low-energy oberservables. We find $M_{GUT}=2.36\times
10^{16}\rm{GeV}$. The dashed lines in the figure show the pure MSSM
running.\label{fig:gaugerunning}}
\end{figure}

This type-II seesaw requires that we have a medium scale for the
mass of the SM triplet Higgs which is present in {\bf 15}-Higgs i.e.
$M_T\sim 10^{14}\rm{GeV}$; this is satisfied if we tune the coupling
$\lambda$ of $\lambda\Phi S\bar{S}$ to $\sim 10^{-2}$ or so since
$M_T\sim \lambda v_U$. Once $\Phi$ get vev and breaks $SU(5)$ to
Standard Model, it also can induce the mass splitting of multiplets
of $S,\bar{S}$. This will affect the unification of coupling. We
display the effect of these mass splittings to the gauge coupling
running as a threshold correction, in Fig.\ref{fig:gaugerunning} and
show that the unification of couplings is maintained and we get a
slight increment in the value of $M_U\simeq 2.36\times 10^{16}$ GeV.

\subsection{{\bf 45} vrs its higher dimensional equivalent}
We also like to comment that a more economical possibility  is to
consider a model that uses a high dimension operator involving with
$\Phi$ instead of the $H$. The matter part of the superpotenial in
this case is given by:
\begin{eqnarray}
W=Y_{15}FFS+Y_{5}TTh+
Y_{\bar{5}}FT\bar{h}+\frac{1}{M_{P}}Y_{24}FT\Phi\bar{h},
\end{eqnarray}
where $M_{P}$ is Planck scale and $H_{24}$ is the $SU(5)$ adjoint
representation used to break $SU(5)$ to $SU(3)\times SU(2)\times
U(1)$. $M_P\sim 10^{19} \rm{GeV}$,vev of $H$ is $\sim
10^{16}\rm{GeV}$ and vev of $\bar{h}$ is $\sim 10^{2}\rm{GeV}$ ,thus
the overall scale of the contribution of this higher dimensional
operator to fermion mass matrices $\sim 100\rm{MeV}$. We have tried
a fitting of data for this model and find it to be unacceptable,
since it gives very large $\sin^2\theta_{23}\sim 0.76-0.8$ which is
around 4-5 $\sigma$.

\subsection{Possible $S_3$ embedding}
An interesting possibility is to embed the $\mu-\tau$ symmetry into
an $S_3$ symmetry. There are two reasons one may consider such an
extension. First is that the neutrino mass matrix of the form that
originates from the ${\bf 15}$ coupling is then given by:

\begin{eqnarray}
{\cal M}_\nu~=~\pmatrix{a & b& b\cr b & a & b\cr b & b& a}
\label{tbm1}
\end{eqnarray}

 which reproduces the interesting tri-bi-maximal mixing
pattern\cite{tbm} which seems to be in very good accord with current
data. SU(5) embedding could then provide corrections to the
tri-bi-maximal mixing. However, mixing matrix obtained from
Eq.\ref{tbm1} is  arbitrary up to a rotation in the 1-3 space due to
the fact that the first and the third  generations are degenerate.
If we add to this matrix the following $\mu-\tau$ symmetric but
$S_3$ breaking matrix of the form:
\begin{eqnarray}
{\cal \delta M}_\nu~=~c\pmatrix{0 & 0 & 0\cr 0 & 1 & -1\cr
0 & -1 & 1}
\end{eqnarray}
This however breaks the $S_3$ symmetry by a large amount.

The second interesting point about the $S_3$ embedding is that
 that in the symmetry limit, the SU(5) model
conserves R-parity automatically, making the dark matter naturally
stable since the term $FFT$ is forbidden by the symmetry.

\section{Conclusion}
In summary, we have discussed the grand unification of apparently
pure leptonic symmetries such as $\mu-\tau$ symmetry into the
quark-lepton unifying Supersymmetric SU(5) model for quarks and
leptons and studied its implications for neutrino mixing angles.
We find that it is possible to have a completely viable SU(5)
model of this type. In this model the neutrino masses arise from a
triplet vev induced type II seesaw mechanism. The presence of
quark lepton unification leads to small deviations from maximal
atmospheric mixing angle and vanishing $\theta_{13}$ implied in
the exact symmetry limit.

 This work is supported by the National Science Foundation grant
no. Phy-0354401

\end{document}